# A New *Lecture-Tutorial* for Teaching about Molecular Excitations and Synchrotron Radiation

**Colin S. Wallace,** University of North Carolina at Chapel Hill, Chapel Hill, NC
**Edward E. Prather,** University of Arizona, Tucson, AZ
**Seth D. Hornstein,** University of Colorado Boulder, Boulder, CO
**Wayne M. Schlingman,** The Ohio State University, Columbus, OH
**Timothy G. Chambers,** University of Michigan, Ann Arbor, MI
**Jack O. Burns,** University of Colorado Boulder, Boulder, CO

Light and spectroscopy are among the most important and frequently taught topics in introductory, college-level, general education astronomy courses (hereafter Astro 101).[1] This is due to the fact that the vast majority of observational data studied by astronomers arrives at Earth in the form of light. While there are many processes by which matter can emit and absorb light, Astro 101 courses typically limit their instruction to the Bohr model of the atom and electron energy level transitions. In this paper, we report on the development of a new *Lecture-Tutorial* to help students learn about other processes that are responsible for the emission and absorption of light, namely molecular rotations, molecular vibrations, and the acceleration of charged particles by magnetic fields. Note that this paper primarily focuses on describing the variety of representations and reasoning tasks designed for this *Lecture-Tutorial*; while the end of this paper highlights some data that is suggestive of the *Lecture-Tutorial*'s effectiveness, our more comprehensive analysis of its efficacy will be presented in a future publication.

The design and development process for this new *Lecture-Tutorial* followed the principles and methodology that guided the development of the original *Lecture-Tutorials for Introductory Astronomy*.[2] Each of the forty-four previously-developed, evidence-based *Lecture-Tutorials* is a two- to six-page worksheet containing carefully sequenced and conceptually-scaffolded Socratic-style questions, utilizing multiple representations of scientific information, that actively engage collaborative student learning groups in constructing more expert-like understandings of a small number of closely-related astrophysical topics. Several prior research studies show that the implementation of *Lecture-Tutorials* can help students significantly increase their understandings of difficult physics and astronomy concepts beyond what they typically achieve during traditional lecture-based instruction.[3]

The development of this *Lecture-Tutorial* was one of the initial projects of a new collaboration between scientists and science education researchers from Associated Universities, Inc. (AUI), the Center for Astronomy Education (CAE) at the University of Arizona (UA), the University of Colorado Boulder (CU-Boulder), and the University of North Carolina at Chapel Hill (UNC). One of the goals of this collaboration is to create the tools STEM instructors need to transform their college classes into active-engagement, collaborative learning environments using foundational and cutting-edge radio astronomy topics. For this particular *Lecture-Tutorial*, our goals were to help students understand:
1) the interactions between light and matter involving molecular rotations, molecular vibrations, and synchrotron radiation;

2) the wavelengths of light and types of spectra that are produced by these different processes;
3) the connection between these processes and real astrophysical phenomena; and
4) the telescopes whose wavelength sensitivities are best suited for observing each particular type of matter-energy interaction.

**A New *Lecture-Tutorial* on Molecular Vibrations, Molecular Rotations, and Synchrotron Radiation**

During the prior research and development on the *Lecture-Tutorial* that focused on the Bohr model of the atom, we found that many students leave a traditional lecture unable to consistently and coherently reason about whether an electron moving from a higher to lower energy level (and vice versa) results in the emission or absorption of light, and how the wavelength of the emitted/absorbed light is related to the energy difference between the electron's initial and final states. We were therefore not surprised to find that students exhibit similar difficulties when asked to reason about how changes in the rotational or vibrational energy state of a molecule are related to the emission or absorption of light. As a result, we begin the new *Lecture-Tutorial* with questions that help students develop their conceptual understandings of, and abilities to reason about, the emission and absorption of different wavelengths of light via molecular vibrations and rotations.

Note that throughout this new *Lecture-Tutorial* we explicitly use many different representations. We hereafter use the term "pedagogical discipline representations" (PDRs) to refer to the body of representations that are useful for teaching and learning about a given discipline, such as astronomy, even though those representations may not be typically used in textbooks or by two experts in a field that are communicating with one another. Different PDRs may have complementary abilities to deepen students' understandings due to the fact that these representations have different scientific and educational affordances.[4] Consequently, when we develop a new *Lecture-Tutorial*, we use the widest variety of PDRs possible, including graphs, equations, pictures, and tables of data, so that we may engage the critical thinking skills of the largest possible range of students and help them become more fluent in the science content they are studying. For example, in the *Lecture-Tutorial* that focuses on the Bohr model of the atom, students reason about a combination of pictorial and word-based representations of photon and atom interactions, electron energy levels, and the relationship between energy level transitions and the wavelength and energy of emitted/absorbed light. The new *Lecture-Tutorial* uses carefully chosen PDRs designed to help students envision the physical behavior of molecules that change energy states due to the emission and absorption of photons, and how these matter energy interactions relate to different representations of spectra. Figures 1-5 show excerpts from the *Lecture-Tutorial* that highlight the range of question types, reasoning tasks, and representations used; the full *Lecture-Tutorial* contains additional questions that aid in the transition between the questions shown in this paper. As shown in Figure 1, we use the number of arcs on either side of a molecule to qualitatively represent its vibrational energy state. More arcs correspond to a vibrational state of greater energy. Figure 2 shows our use of arrows to qualitatively represent the rotational energy state of a molecule. Longer arrows correspond to a rotational state of greater energy. Through field tests with over 1000 students, we found that these arrows and arcs facilitate students' discourse about changes in rotational and vibrational energy states, and, as a result, students are able, post-instruction, to incorporate these pictorial representations into explanations that are more scientifically coherent than those provided after traditional lecture-based instruction.

[Figure 1 about here.]
[Figure 2 about here.]

Figures 1 and 2 also show a sequence of questions used in the new *Lecture-Tutorial* to engage students about how changes in a molecule's vibrational or rotational energy state corresponds to the emission or absorption of a photon. The new *Lecture-Tutorial* requires students to reason about how the energy/wavelength/frequency of a photon that is emitted or absorbed as a result of one molecular energy state transition compares to the energy/wavelength/frequency of a photon emitted or absorbed due to a different energy state transition. For example, students are expected to realize that both Case B and C in Figure 1 correspond to the absorption of a photon, although the photon absorbed in Case C must have a longer wavelength (or lower energy) than the photon absorbed in Case B. Reasoning tasks associated with the energy/wavelength/frequency of an emitted/absorbed photon target and reinforce conceptual ideas and reasoning pathways that have been developed in other *Lecture-Tutorials*.

Section II of the new *Lecture-Tutorial* is devoted to synchrotron radiation. It focuses on helping students understand that synchrotron radiation has a continuous spectrum that is not dependent on temperature (*i.e.*, it is non-thermal). Figure 3 shows a question from this section asking students to sketch a qualitatively correct spectrum for the light produced by a pulsar, whose surface emits a thermal radiation and whose jets emit synchrotron radiation.

[Figure 3 about here.]

Section III requires students to synthesize all the ideas covered in Sections I and II and apply them to real astrophysical contexts. Figure 4 shows a question from this section that asks students to select the appropriate telescope needed to make a real astronomical observation. To correctly answer this question, a student must recognize that the rotating organic molecules will emit in the radio and produce a line spectrum. The Atacama Large Millimeter Array (ALMA) is the only radio telescope listed, so it is the only one out of the four possible telescopes that can be used to observe this spectrum. Figure 5 shows the *Lecture-Tutorial*'s culminating questions, which require students to evaluate multiple linked representations and to use their knowledge of molecular vibrations, molecular rotations, and synchrotron radiation to re-represent all incorrect information. To complete this task, students must recognize that 1) the vibrational energy state transition depicted would lead to the emission, not absorption, of a photon, 2) the photon would not be in the radio, since vibrational energy state transitions generally produce infrared light, and 3) the spectrum shown is incorrect, since it depicts the continuous spectrum of synchrotron radiation and not the line spectrum of vibrational energy state transitions.

[Figure 4 about here.]
[Figure 5 about here.]

**Assessment**

In order to gain insight into the effectiveness of this new *Lecture-Tutorial*, we created a survey of twelve conceptually challenging questions. These questions probe the same content as the *Lecture-Tutorial*, although they differ in their context and/or format from the questions asked on the *Lecture-Tutorial* itself. Figure 6 shows a subset of these questions. The assessment's face validity and content accuracy were established by the review of content experts in radio astronomy.

[Figure 6 about here.]

We pilot tested this new *Lecture-Tutorial* with over 1000 students over three semesters at UA and CU-Boulder. Over these three semesters, we designed many different assessment methodologies in order to measure the effects of different implementations of this *Lecture-Tutorial*.

For this publication, we report the data we gathered from only one of the many methodologies for assessing the efficacy of the *Lecture-Tutorial*; results from our larger dataset will appear in a future publication. In the spring of 2014, we tested the implementation of *Lecture-Tutorial* by examining the understandings of a population of students enrolled in two sections of Astro 101 at CU-Boulder. Both sections had similar demographics and distributions of overall grades. The two sections each received a carefully crafted lecture from the same instructor on the relevant astrophysics of each concept before doing the *Lecture-Tutorial*. One section of this Astro 101 course took the twelve-question assessment before any instruction on the topics of the *Lecture-Tutorial* (pre-test). This section of the course also took the twelve-question assessment after the lecture (post-lecture) but before they worked through the *Lecture-Tutorial* activity (pre-*Lecture-Tutorial*). The other section did not take a pre-test but did take the assessment after completing the *Lecture-Tutorial* activity (post-*Lecture-Tutorial*); note that the students in this section were also asked two low-level Think-Pair-Share questions during lecture. The pre-*Lecture-Tutorial* and post-*Lecture-Tutorial* results from these two Astro 101 sections were each compared to the pre-test results. This research design allows us to isolate the improvement in understanding achieved from lecture alone and compare this result with the amount of gain in understanding achieved when the active engagement *Lecture-Tutorial* is done in class immediately following lecture.

The single section that took the twelve-question assessment as a pre-test had an average pre-test score of 22% correct ($n = 112$). This pre-test score is equal to the other pre-test scores from our larger dataset, in which Astro 101 populations at multiple institutions who received no prior instruction on these topics, also had an average score on this assessment of 22% correct. We therefore assume a pre-test score of 22% for the other spring 2014 Astro 101 section at CU-Boulder, for which we did not administer the pre-test. By not administering the pre-test to this population, we also ensure that the students who took the assessment post-*Lecture-Tutorial* had never seen any of the twelve conceptual questions used in this study (this is not true for the post-lecture section in our comparison below).

The average correct score for the students post-lecture was 49% ($n = 73$). The average correct score for the students post-*Lecture-Tutorial* was 57% ($n = 50$). These results suggest that a well-focused lecture may indeed help students improve their understanding of these topics (pre- to post-lecture gain of 27%). Additionally, through the implementation of this new *Lecture-Tutorial*, which takes only fifteen minutes (approximately) of class time, a class may experience an additional increase in understanding of nearly an entire letter grade (post-lecture to post-*Lecture-Tutorial* gain of 8%). This result is consistent with the gains in students' understandings on other astrophysics topics achieved by incorporating other *Lecture-Tutorials* into Astro 101 courses.[3] A z-test comparing the post-lecture and post-*Lecture-Tutorial* means, and a Student's t-test comparing the post-lecture and post-*Lecture-Tutorial* distributions both result in *p*-values of $p = 0.06$. This suggests that the differences between the post-lecture and post-*Lecture-Tutorial* populations would be produced by chance only 6% of the time.

Further evidence for the effectiveness of the *Lecture-Tutorial* can be seen in Figure 7, which shows the percentage of students in both populations in bins based on the percentage correct on the post-test. Note that the post-lecture distribution has a greater concentration of scores at the lower end of the distribution (a range of scores that also happens to include the pre-instruction average) than does the post-*Lecture-Tutorial* group. Overall, this analysis suggests that the new *Lecture-Tutorial* may be effective at moving a subset of students to a higher-level of understanding, which they may not have achieved after lecture alone. This result is consistent with the levels of understanding students have been shown to achieve after completing other *Lecture-Tutorials*; as noted above, we will continue to investigate the effectiveness of different implementations of this *Lecture-Tutorial* in future publications.

**Availability of the *Lecture-Tutorial***

Instructors who wish to obtain a copy of the current version of the *Lecture-Tutorial*, as well as the accompanying lecture slides and Think-Pair-Share ("clicker") questions that have been developed to support this *Lecture-Tutorial*, should contact the authors. A copy of the *Lecture-Tutorial* is also included as an on-line appendix.

**Acknowledgements**

This work was made possible by the generous support of Associated Universities, Inc. (AUI). We would particularly like to thank Ethan Schreier, John Mester, John Stoke, and Timothy Spuck for their guidance and support. Additionally, we would like to thank Johanna Teske and Jeffrey Eckenrode for their help with the pilot testing and data collection of this curriculum.

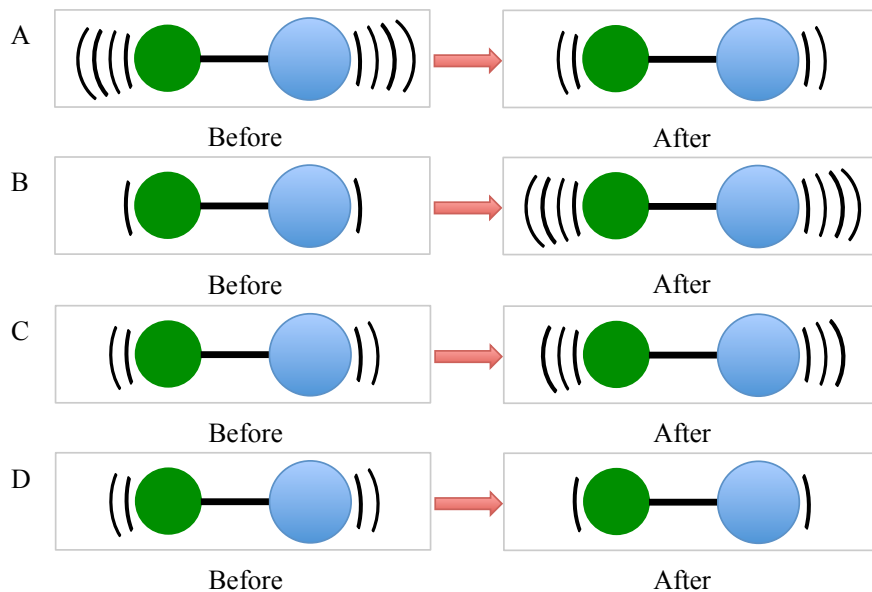

A molecule can change its vibrational energy state by emitting or absorbing an infrared photon. In each of the four cases shown below (A-D), a molecule is shown transitioning between two different energy states depending on whether the molecule absorbed or emitted a photon.

1) Which of the cases (A-D) shown above correspond with the absorption of light and which correspond with the emission of light? Draw in a squiggly arrow representing a photon with the appropriate direction for each case (A-D).

2) In which case was a photon with the longest wavelength absorbed? Explain your reasoning.

3) In which case was a photon with the greatest energy emitted? Explain your reasoning.

**Fig. 1. Sample questions from the *Lecture-Tutorial* that help develop students' understandings of the emission and absorption of light by molecular vibrations.**

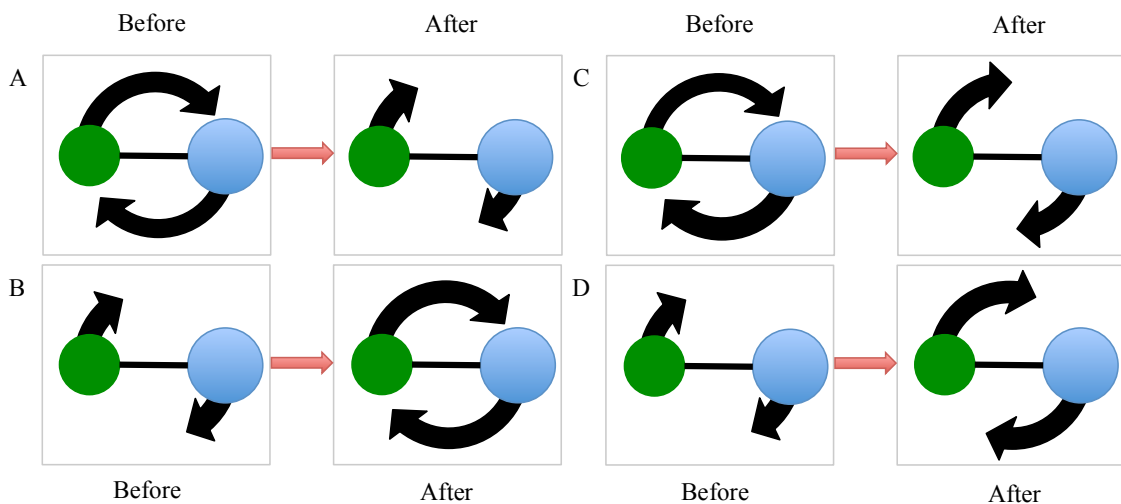

A molecule can change its rotational energy state by emitting or absorbing a radio photon. In each of the four cases shown below (A-D), a molecule is shown transitioning between two different energy states depending on whether the molecule absorbed or emitted a photon.

4) Which of the cases (A-D) shown above correspond with the absorption of light and which correspond with the emission of light? Draw in a squiggly arrow representing a photon with the appropriate direction for each case (A-D).

5) In which case was a photon with the shortest wavelength emitted? Explain your reasoning.

6) In which case was a photon with the least energy is absorbed? Explain your reasoning.

**Fig. 2.** Sample questions from the *Lecture-Tutorial* that help develop students' understandings of the emission and absorption of light by molecular rotations.

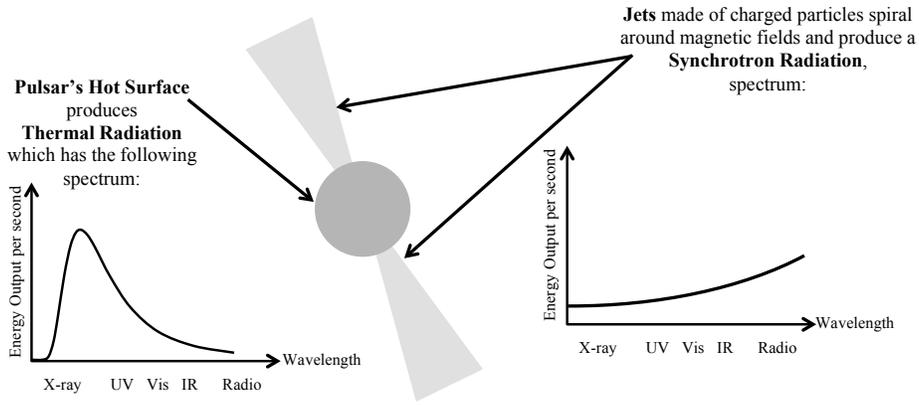

11) Imagine you could observe the pulsar at all wavelengths of light, from X-rays to radio. On the graph below, sketch the spectrum of light you would detect that results from the **combination** of the synchrotron radiation from the pulsar's jets and the thermal radiation from the pulsar's surface.

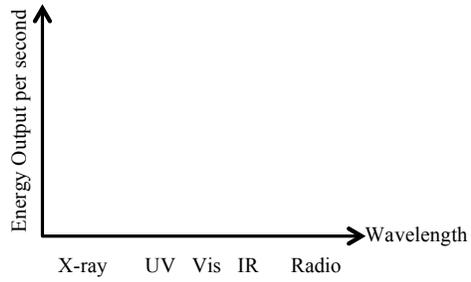

**Fig. 3. Sample question from the *Lecture-Tutorial* that asks students to reason about a combined thermal and synchrotron spectrum.**

Here are four real telescopes:
- The Atacama Large Millimeter Array (ALMA) is a radio telescope located in Chile
- The Chandra X-ray Observatory is an X-ray telescope located in space, orbiting the Earth
- The Spitzer Space Telescope is an IR telescope located in space, orbiting the Sun
- The Galaxy Evolution Explorer (GALEX) is a UV telescope located in space, orbiting the Earth

Use the above information to help answer the following questions.

14) Astronomers, trying to understand the origins of life, observe rotating organic molecules in nebulae.
   a) Rotating organic molecules produce a(n) _______ (**circle one of the following three choices:** *synchrotron / thermal / emission or absorption line*) spectrum.

   b) Which telescope listed above would be best to detect the light from rotating organic molecules in nebulae? Explain your reasoning.

15) Here is a spectrum of light:

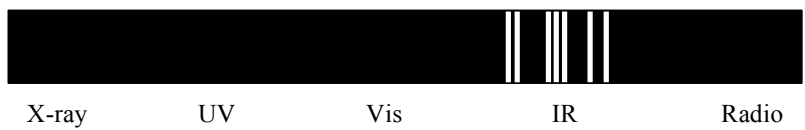

   X-ray     UV     Vis     IR     Radio

   a) This spectrum is produced by ____ (**circle one of the following five choices**).
      - the hot surface of a pulsar
      - charged particles spiraling around a magnetic field
      - electrons changing energy states
      - molecules changing vibrational states
      - molecules changing rotational states

   b) Which telescope listed above would you use to detect this spectrum?

**Fig. 4. Sample question from the *Lecture-Tutorial* that requires students to select the appropriate telescope based on what they have learned from earlier sections of the *Lecture-Tutorial*.**

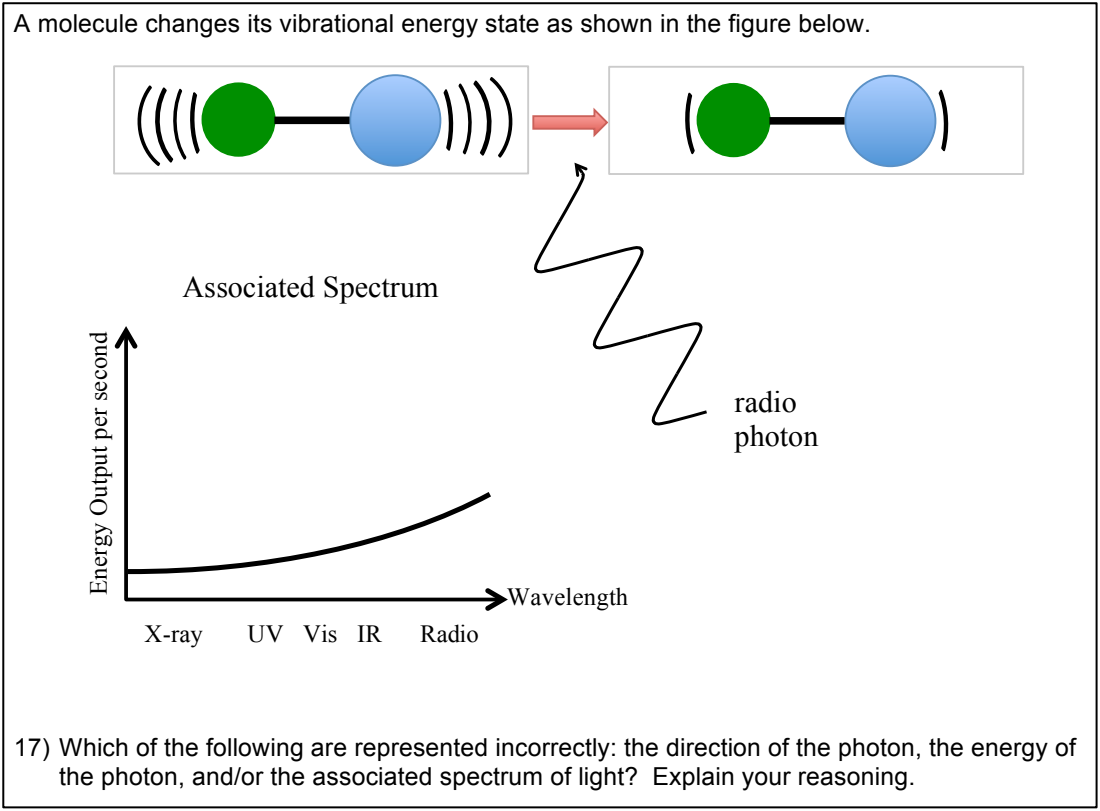

Fig. 5. Sample question from the *Lecture-Tutorial* that requires students to identify each problem with the above picture.

1) Which drawing corresponds to the emission of a photon with the least energy?
    a. A
    b. B
    c. C
    d. D
    e. There's not enough information to tell.

2) Which one of the following telescopes would you use if you wanted to detect the greatest amount of radiation by charged particles spiraling around the magnetic field of a black hole?
    a. the infrared Spitzer Space Telescope
    b. the Very Large Array radio telescope
    c. the Chandra X-ray Observatory
    d. the GALEX ultraviolet telescope

3) Which of the following describes the most likely change a molecule experiences after it absorbed an infrared photon?
    a. The molecule rotates faster
    b. The molecule rotates slower
    c. The molecule vibrates faster
    d. The molecule vibrates slower

**Fig. 6. Sample questions assessment used to measure the efficacy of the *Lecture-Tutorial*.**

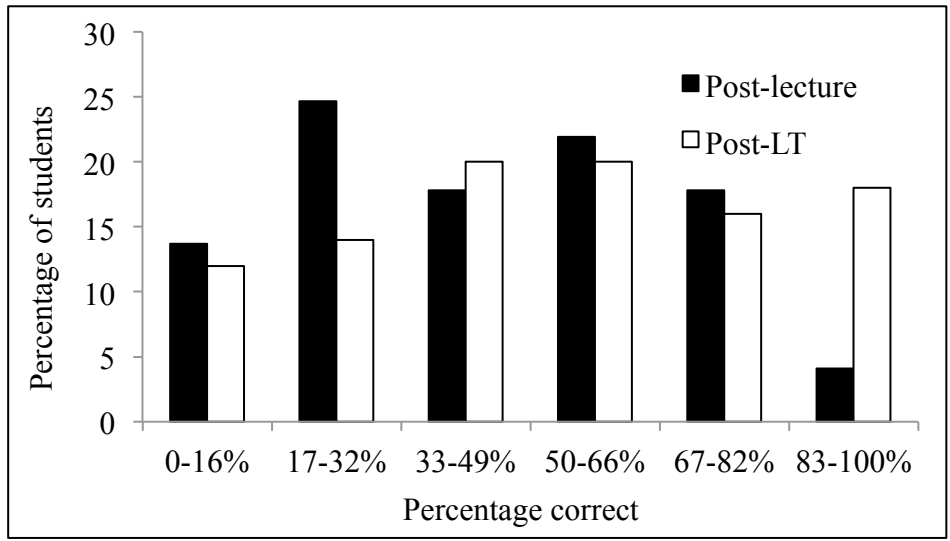

**Fig. 7. Histograms of the post-lecture and post-*Lecture-Tutorial* populations binned by percentage correct on the post-test.**